\documentclass[12pt,a4paper]{article}

\usepackage[latin1]{inputenc}
\usepackage{amsmath}
\usepackage{amsfonts}
\usepackage{amssymb} 
\usepackage{graphicx}
\usepackage{a4wide}
\usepackage{url}
\usepackage{slashbox}
\usepackage{supertabular}

\author{H.~M\scshape {outarde}\footnote{herve.moutarde@cea.fr} \\ 
	{\small \textit{CEA, Centre de Saclay, IRFU/Service de Physique Nucléaire}} \\
	{\small \textit{F-91191 Gif-sur-Yvette, France}}
}

\title{\textbf{Extraction of the Compton Form Factor $\mathcal{H}$ from DVCS measurements at Jefferson Lab}}

\date{}

\newcommand{\ie}{\textit{i.e. }}

\newcommand{\etal}{\textit{et al. }}
\newcommand{\refeq}[1]{Eq.~(\ref{#1})}
\newcommand{\reftab}[1]{Tab.~\ref{#1}}
\newcommand{\reffig}[1]{Fig.~\ref{#1}}

\newcommand{\xb}{x_B}
\newcommand{\gevcarre}{\textrm{GeV}^2}

\newcommand{\ImH}{Im \mathcal{H}}
\newcommand{\ReH}{Re \mathcal{H}}
\newcommand{\ImE}{Im \mathcal{E}}

\newcommand{\ImHT}{Im \mathcal{\tilde{H}}}

\newcommand{\ImF}{Im \mathcal{F}}
\newcommand{\ReF}{Re \mathcal{F}}
\newcommand{\ImFT}{Im \mathcal{\tilde{F}}}
\newcommand{\ReFT}{Re \mathcal{\tilde{F}}}

\begin{document}

\maketitle

\begin{abstract}
In the framework of Generalised Parton Distributions, we study  the helicity-dependent and independent cross sections measured in Hall A and the beam spin asymmetries measured in Hall B at Jefferson Laboratory. We perform a global fit of these data and fits on each kinematical bin. We extract the real and imaginary parts of the Compton Form Factor $\mathcal{H}$ under the main hypothesis of dominance of the Generalised Parton Distribution $H$ and twist 2 accuracy. We discuss our results and compare to previous extractions as well as to the VGG model. We pay extra attention to the estimation of errors on the extraction of $\mathcal{H}$. 
\end{abstract}

%%%%%%%%%%%%%%%%%%%%%%%%%%%%%%%%%%%%%%%%%%%%%%%%%%%%%%%%%%%%%%%%%%%%%%%%%%%%%%%%%%%%%%%%%%%%%%%%%%%%
%%%%%%%%%%                                  Introduction                                  %%%%%%%%%%
%%%%%%%%%%%%%%%%%%%%%%%%%%%%%%%%%%%%%%%%%%%%%%%%%%%%%%%%%%%%%%%%%%%%%%%%%%%%%%%%%%%%%%%%%%%%%%%%%%%%

\section*{Introduction}

Since it was realised that Generalised Parton Distributions (GPD) were reachable through an harmonic analysis of the Deeply Virtual Compton Scattering (DVCS) process \cite{Ji96}, \cite{DGPR97}, the study of the connection between GPDs and DVCS has been a very active field of research, concerning theoretical developments (see the reviews \cite{GPV01}, \cite{Die03}, \cite{BR05} and \cite{BP07}) as well as experimental ones (\cite{Air01}, \cite{Adl01}, \cite{Ste01}, \cite{Che03}, \cite{Che06}, \cite{Mun06}, \cite{Aar07}, \cite{Gir07} and \cite{Air08}). After the first dedicated experiments and at the beginning of the experimental GPD program, it is already worth trying to extract GPDs from measurements. 

The present work addresses this question, and illustrates it with recent JLab data, namely beam spin asymmetries (BSA) \cite{Gir07} and helicity-dependent and independent cross sections \cite{Mun06}. These data offer a large kinematic coverage and a fine kinematic binning, which are interesting features for our purpose. However, the methods we use and the conclusions we come to are presumably not restricted to JLab kinematics, and may be of interest for other experimental set-up. Furthermore, this is one of the first global fits of these measurements, and such fits are necessary to the completion of the experimental GPD program.

The harmonic analysis of $ep \rightarrow ep\gamma$ cross-sections has so far relied on the 2002 work of A.V.~Belitsky, D.~Müller and A.~Kirchner \cite{BMK02}. In this formalism, the interference between the Bethe-Heitler (BH) and DVCS processes was treated with a leading order approximation of the BH part. This assumption was removed by A.V.~Belitsky and D.~Müller in \cite{BM08} in the case of a spinless target, and by P.A.M.~Guichon and M.~Vanderhaegen in the case of a proton target \cite{GV08}. In all the following, we will use the expressions from the latter.

The first section of this paper describes JLab data, the difference of the evaluation of the $ep \rightarrow ep\gamma$ cross sections by \cite{BMK02} and \cite{GV08}, and outlines our hypothesis. In the second part we explain our fitting strategy, which is twofold~: we perform a global fit of the selected JLab data and fits on each kinematic bin (which we will be refering to as \emph{local fits}). In the third section we systematically compare the results of global and local fits, and keep the output of the global fit as our best solution. We then discuss the extracted values of $\mathcal{H}$ and compare them to previous studies.

%%%%%%%%%%%%%%%%%%%%%%%%%%%%%%%%%%%%%%%%%%%%%%%%%%%%%%%%%%%%%%%%%%%%%%%%%%%%%%%%%%%%%%%%%%%%%%%%%%%%
%%%%%%%%%%                                 Première partie                                %%%%%%%%%%
%%%%%%%%%%%%%%%%%%%%%%%%%%%%%%%%%%%%%%%%%%%%%%%%%%%%%%%%%%%%%%%%%%%%%%%%%%%%%%%%%%%%%%%%%%%%%%%%%%%%

\section{Preliminary analysis}

In this study, $\xb$ is the standard Bjorken variable, $Q^2$ the virtuality of the initial photon and $t$ the square momentum transfer between initial and final protons.

\subsection{Description of the selected JLab data}

C.~Mu{\~n}oz-Camacho \etal \cite{Mun06} published helicity-dependent and helicity-independent cross sections for $\xb$ = 0.36 and $t$ between -0.33 and -0.17~$\gevcarre$. Helicity-independent cross sections are given at $Q^2$ = 2.3~$\gevcarre$ only, while helicity-dependent cross sections are measured for $Q^2$ between 1.5 and 2.3~$\gevcarre$. Data come in 12 bins (3 values of $Q^2$ and 4 values of $t$). Helicity-dependent cross sections are given with relatively large uncertainties (from 20~\% to 100~\%, depending on the bins and on the value of the angle between leptonic and hadronic planes) while helicity-independent cross sections are usually accurate at the 5~\% level.

F.-X.~Girod \etal \cite{Gir07} released BSAs over a wide kinematic range ($\xb$ from 0.11 to 0.58, $Q^2$ from 1 to 4.8 $\gevcarre$ and $t$ from -1.8 to -0.09 $\gevcarre$, described by 62 multi-dimensional bins (5 values of $t$, 5 values of $\xb$ and 4 values of $Q^2$). Their accuracy is 25~\% on average, ranging from 5 to 100~\% depending on the bins and on the value of the angle between leptonic and hadronic planes.

These two sets of measurements have kinematic configurations in common, allowing us to perform cross-checks by evaluating a BSA through the ratio of helicity-dependent to helicity-independent cross sections. This simple exercise shows no unexpected discrepancy between Halls A and B measurements. This implies that both sets of data are consistent and can be used in a global fit. %This global fit will thus be constrained by physical quantities with both relative and absolute meanings.

\subsection{DVCS at leading twist}

Four GPDs $H$, $E$, $\tilde{H}$ and $\tilde{E}$ appear at twist 2, but the cross sections depend on the Compton Form Factors (CFF). The convention of \cite{BMK02} is used to define the CFFs~:
\begin{eqnarray}
	\mathcal{F} & = & \int_{-1}^{+1} dx \, F(x,\xi,t) \left( \frac{1}{\xi-x-i\epsilon} - \frac{1}{\xi+x-i\epsilon} \right) \quad (F = H \textrm{ or } E) \\
	\mathcal{\tilde{F}} & = & \int_{-1}^{+1} dx \, \tilde{F}(x,\xi,t) \left( \frac{1}{\xi-x-i\epsilon} + \frac{1}{\xi+x-i\epsilon} \right) \quad (\tilde{F} = \tilde{H} \textrm{ or } \tilde{E})
\end{eqnarray}
and $\xi = \xb \frac{1+\frac{t}{2 Q^2}}{2-\xb+\frac{\xb t}{Q^2}}$ is the generalised Bjorken variable \cite{BMK02}, \cite{GV08}. The complex integration kernel yields a real and an imaginary part to the CFFs~:
\begin{eqnarray}
	\ReF & = & \mathcal{P} \int_{-1}^{+1} dx \, F(x,\xi,t) \left( \frac{1}{\xi-x} - \frac{1}{\xi+x} \right) \quad (F = H \textrm{ or } E) \label{EqRealPartHE} \\
	\ImF & = & \pi \Big( F(\xi,\xi,t) - F(-\xi,\xi,t) \Big)  \quad (F = H \textrm{ or } E)  \label{EqImaginaryPartHE} 
\end{eqnarray} 
and
\begin{eqnarray}
	\ReFT & = & \mathcal{P} \int_{-1}^{+1} dx \, F(x,\xi,t) \left( \frac{1}{\xi-x} - \frac{1}{\xi+x} \right) \quad (\tilde{F} = \tilde{H} \textrm{ or } \tilde{E}) \\
	\ImFT & = & \pi \Big( F(\xi,\xi,t) + F(-\xi,\xi,t) \Big)  \quad (\tilde{F} = \tilde{H} \textrm{ or } \tilde{E})
\end{eqnarray} 
where the symbol $\mathcal{P}$ denotes the principal value of the integral.

The reak and imaginary parts of a CFF are related by dispersion relations due to analycity properties (see \cite{Ter05}, \cite{AT07} or \cite{DI07}). However the unknown substraction (the D-term \cite{PW99}) and the limited kinematic range of our data make this constraint rather weak. In this work we consider the real and imaginary parts as independent (local fits) except in the case where we use an explicit parametrisation of the GPD $H$ to compute the CFF (global fit).

\subsection{The formalism of P.A.M.~Guichon and M.~Vanderhaeghen and its consequences}

The recent work of P.A.M.~Guichon and M.~Vanderhaeghen (GV) gives analytic expression for the $ep \rightarrow ep\gamma$ cross section~; these formulae are embodied in a Mathematica package \cite{GV08}, which we used to build a C++/ROOT library. 

The $ep \rightarrow ep\gamma$ cross section is classicaly divided into three parts, namely BH, VCS and interference between the BH and VCS cross sections. The contribution of the BH amplitude to the interference is treated exactly. This introduces some differences with respect to the well-known BMK expressions and two new important qualitative features to the discussion of the measurements under scrutiny. Let us mention before an important point for actual computation of cross sections~: even if the GV expressions rely on the same tensorial decomposition of the hadronic VCS tensor than BMK expressions, the GV kinematic conventions follow those used in early versions in the VGG code detailed in \cite{VG98}. The BMK and GV formalisms thus use different definitions for the angle $\phi$ between the hadronic and leptonic planes, and these definitions do not follow the Trento convention \cite{Bac04} chosen to describe the measurements~: $\phi_{{\rm {\tiny Trento}}} = \pi - \phi_{{\rm {\tiny BMK}}}$ and $\phi_{{\rm {\tiny Trento}}} = 2 \pi - \phi$.

Concerning Hall A measurements, the BMK formalism restricted to twist 2 asserts that helicity-dependent cross sections write~:
\begin{equation}
	\frac{1}{2} \left[ \frac{d^4\sigma^+}{d^4\Phi} - \frac{d^4\sigma^-}{d^4\Phi} \right] = C_1 \  \sin \ \phi \quad Im \ \Big( \mathcal{H} + \frac{\xb}{2-\xb} ( 1 + \frac{F_2}{F_1} ) \mathcal{\tilde{H}} - \frac{t}{4 M^2} \frac{F_2}{F_1} \mathcal{E} \Big)
\end{equation}
where $d^4\Phi = dQ^2 d\xb dt d\phi$, $F_1$ and $F_2$ are the Dirac and Pauli form factor, $M$ the proton mass and $C_1$ is a constant irrelevant for our purpose. Since the $Q^2$-dependence of this cross section is factorised, this expression allows a study of scaling without having to disentangle the different CFFs as in \cite{Mun06}.

With the exact equations of \cite{GV08}, an helicity-dependent cross section does not have such a simple form. It writes~:
\begin{equation}
\label{EqInterfGV}
	\frac{1}{2} \left[ \frac{d^4\sigma^+}{d^4\Phi} - \frac{d^4\sigma^-}{d^4\Phi} \right] = C_2 	\ \sin \phi \quad Im \Big( \mathcal{H} + c_\mathcal{E} \ \mathcal{E} + c_\mathcal{\tilde{H}} \ \mathcal{\tilde{H}} + c_\mathcal{\tilde{E}} \ \mathcal{\tilde{E}} \Big) + \ldots
\end{equation}
where $C_2$ is a constant irrelevant for our purpose. The dots stand for power-suppressed contributions. The test of scaling is more involved now since the coefficients $c_\mathcal{E}$, $c_\mathcal{\tilde{H}}$ and $c_\mathcal{\tilde{E}}$ do depend on $Q^2$. At given $\xb$ and $t$, the coefficients of \refeq{EqInterfGV} $c_\mathcal{E}$, $c_\mathcal{\tilde{H}}$ and $c_\mathcal{\tilde{E}}$ vary respectively by 20~$\%$, 6~$\%$ and 38~$\%$. If we only fit the combination of CFFs appearing in $Im(\ldots)$ in \refeq{EqInterfGV}, the kinematic $Q^2$-dependence of $c_\mathcal{E}$, $c_\mathcal{\tilde{H}}$ and $c_\mathcal{\tilde{E}}$ may appear as a scaling deviation of the same magnitude.

Differences also arise in the expression of a BSA. Its dependence on the angle $\phi$ between the leptonic and hadronic planes takes the following form~:
\begin{equation}
\label{EqBSAGV}
	\textrm{BSA} = \frac{a \sin \phi + b \sin 2 \phi}{1 + c \cos \phi + d \cos 2 \phi + e \cos 3 \phi}
\end{equation}
where $a = \mathcal{O}(Q^{-1})$, $b = \mathcal{O}(Q^{-4})$, $c = \mathcal{O}(Q^{-1})$ $d = \mathcal{O}(Q^{-2})$ and $e = \mathcal{O}(Q^{-5})$ are real numbers. In the BMK picture, the coefficients $b$, $d$ and $e$ are higher-twist contributions. As a straightforward consequence, we see that the 90$^\circ$ asymmetry is no longer proportional to the imaginary part of a linear combination of CFFs. 

Moreover the coefficient $c$ of \refeq{EqBSAGV} now depends on the imaginary part of CFFs, and not only on the real part as in \cite{BMK02}. This precludes a clean separation of the real and imaginary parts of 
\begin{displaymath}
F_1 \mathcal{H} + \frac{\xb}{2-\xb} ( F_1 + F_2 ) \mathcal{\tilde{H}} - \frac{t}{4 M^2} F_2 \mathcal{E}
\end{displaymath} 
through the distinct measurements of the helicity-dependent and helicity-independent cross sections as was done previously \cite{Mun06}.

\subsection{Main assumptions}

P.A.M.~Guichon and M.~Vanderhaeghen provide exact analytical expressions of the $ep \rightarrow ep\gamma$ observables as functions of the DVCS amplitudes. The latter can be written in terms of GPDs through the usual twist expansion \cite{Ji96}. In this work we restrict ourselves to the twist 2 approximation. This is a reasonable assumption, since C.~Mu{\~n}oz-Camacho \etal \cite{Mun06} claimed the observation of early $Q^2$-scaling.

As we are considering experiments on a proton target, we neglect $E$, $\tilde{H}$ and $\tilde{E}$ ($H$-dominance) for the following reasons. Firstly, this is supported by kinematics~: in \refeq{EqInterfGV}, the coefficient $c_\mathcal{E}$ varies between 0.05 and 0.28, $c_\mathcal{\tilde{H}}$ between 0.3 and 1., and $c_\mathcal{\tilde{E}}$ between 2.~10$^{-5}$ and 0.015 for the 52 kinematic configurations of Hall B data satisfying $\frac{|t|}{Q^2} < \frac{1}{2}$. Secondly, for small $t$ and $\xi$, we expect $\frac{\tilde{H}}{H}$ to be close to $\frac{\Delta q}{q}$ \ie $\frac{1}{4}$. Thirdly, we can check, for instance thanks to the VGG model \cite{VG98}, \cite{VGG98}, \cite{VGG99} and \cite{GPRV05}, that the relative sizes of $\ImE$ and $\ImHT$ to $\ImH$ are similar~: $\frac{\ImE}{\ImH}$ varies between 0.21 and 0.92, and $\frac{\ImHT}{\ImH}$ between 0.13 and 0.91 for the same set of 52 Hall B kinematic configurations. This indicates that the hierarchy between the kinematic coefficients reflects the hierarchy of contributions to the interference. 

Thus assuming $H$-dominance, we may hope to extract information on $H$ from BSAs or helicity-dependent cross sections with a systematic error of 20~\% to 50~\%\footnote{A direct test of the $H$-dominance assumption with the VGG model gives an upper bound of 25~\%. This is comparable to the typical statistical uncertainty on BSAs.}, this approximation being better at small $t$. The advantage of this approach is the dramatic decrease of the number of degrees of freedom involved in fits. M.~Guidal indeed showed in a recent work \cite{Gui08} in the same kinematic region that it is not possible to extract sensible information about the real and imaginary parts of $\mathcal{H}$, $\mathcal{E}$, $\tilde{\mathcal{H}}$ and $\tilde{\mathcal{E}}$ by direct fits of helicity-dependent and independent cross sections. More specifically, keeping only the dominant coefficients in \refeq{EqBSAGV} gives the minimal functional form~:
\begin{equation}
\label{EqBSAGVReduced}
	\textrm{BSA} \simeq \frac{a \sin \phi}{1 + c \cos \phi}
\end{equation}
A direct fit of BSAs to this reduced expression on each ($\xb$,$Q^2$,$t$)-bin, along the lines of \cite{Gir07}, shows that the coefficient $c$ is compatible with 0 (while with a marked trend to negative values) within error bars for 25 bins over the 52 bins for which $\frac{|t|}{Q^2}$ is less than $\frac{1}{2}$. Extracting the real part of the CFFs contained in the coefficient $c$ of \refeq{EqBSAGVReduced} is thus a demanding task.

%%%%%%%%%%%%%%%%%%%%%%%%%%%%%%%%%%%%%%%%%%%%%%%%%%%%%%%%%%%%%%%%%%%%%%%%%%%%%%%%%%%%%%%%%%%%%%%%%%%%
%%%%%%%%%%                                 Deuxième partie                                %%%%%%%%%%
%%%%%%%%%%%%%%%%%%%%%%%%%%%%%%%%%%%%%%%%%%%%%%%%%%%%%%%%%%%%%%%%%%%%%%%%%%%%%%%%%%%%%%%%%%%%%%%%%%%%

\section{Fitting strategies}

The possibility to study GPDs in DVCS rests on factorisation theorems \cite{CF98}, which require a small value of $\frac{|t|}{Q^2}$. In the following, we restrict ourselves to kinematic configurations for which $\frac{|t|}{Q^2} < \frac{1}{2}$.

\subsection{Local fits}

There are at most 12 $\phi$-bins in each Hall B ($\xb$,$Q^2$,$t$)-bin, and 24 $\phi$-bins in each Hall A ($\xb$,$Q^2$,$t$)-bin. $\ReH$ and $\ImH$ are the free parameters of the fits.

We estimate the systematic errors associated to our $H$-dominance hypothesis by first fitting data setting the subdominant GPDs to 0, then fitting the same data setting the subdominant GPDs to their VGG value, and computing the difference.

\subsection{Global fits}

Turning to global fits will help to decrease the statistical uncertainties on the fitted parameters. Moreover, we will benefit from the wide kinematic coverage of Hall B data, and of the accuracy of Hall A measurements in the same fit.

Since we are interested only in extracting values of CFFs, we will not try to extrapolate outside the kinematic region of the measurements we consider. This allows us to use a polynomial parametrisation to perform the fits. The forthcoming difficulty will be the evaluation of the systematic uncertainty related to that phenomenological choice.

\subsubsection{A parametrisation of $H$ from the dual model}

The singlet combinations $H_+$ is~:
\begin{equation}
	H_+(x,\xi,t,Q^2) = H(x,\xi,t,Q^2) - H(-x,\xi,t,Q^2)
	\label{eqdefHElecPlusHMagPlus}
\end{equation}
This is the quantity which is accessible through DVCS.

In the framework of the dual model for a spin $\frac{1}{2}$ target, and assuming $H$-dominance, $H_+$ can be formally expanded according to \cite{PSTS08}~:
\begin{equation}
	H_+(x,\xi,t,Q^2) = 2 \sum_{n=0}^\infty \sum_{l=0}^{n+1} B_{nl}(t,Q^2) \theta \left( 1 - \frac{x^2}{\xi^2} \right) \ \left( 1 - \frac{x^2}{\xi^2} \right) \ C_{2n+1}^\frac{3}{2}\left( \frac{x}{\xi} \right) \ P_{2l}\left( \frac{1}{\xi} \right)
	\label{eqseriesformelle}
\end{equation}
This formal expansion can be resummed as a Gegenbauer polynomial expansion (\cite{PS02}, \cite{GT06}, \cite{GT08} and \cite{KM99})~:
\begin{equation}
	H_+(x,\xi,t,Q^2) = 2 ( 1 - x^2 ) \sum_{n=0}^\infty A_n(\xi,t,Q^2) C_{2n+1}^{3/2}(x)
	\label{eqdefHElecPlusGegenbauer} \\
\end{equation}
The coefficients $A_n$ are defined by~:
\begin{equation}
	A_n(\xi,t,Q^2) = - \frac{4n+5}{(n+1)(2n+3)} \sum_{p=0}^n \xi R_{np}(\xi) \frac{(p+1)(2p+3)}{4p+5} \sum_{l=0}^{p+1} B_{pl}(t,Q^2) P_{2l}\left( \frac{1}{\xi}\right)
	\label{eqdefAElec}
\end{equation}
where $P_{2l}$ is a Legendre polynomial and $R_{np}(\xi)$ is a polynomial the degree of which is $2n+1$~:
\begin{equation}
\label{eqdfrnp}
R_{np}(\xi) = (-1)^{(n+p+1)} \frac{\Gamma\left(\frac{5}{2}+n+p\right)}{\Gamma(n-p+1) \Gamma\left(\frac{5}{2}+2p\right)} \xi^{(2p+1)} \vphantom{.}_2F_1\left(p-n,\frac{5}{2}+n+p,\frac{7}{2}+2p,\xi^2\right)
\end{equation}
with $\vphantom{.}_2F_1$ the Gauss hypergeometric function.

The $Q^2$-evolution of $B_{pl}(t,Q^2)$ is given at leading order in \cite{GT06}~:
\begin{equation}
\label{eqrunningq2}
B_{pl}(t,Q^2) = B_{pl}(t,Q_0^2) \left( \frac{\ln \frac{Q_0^2}{\Lambda^2}}{\ln \frac{Q^2}{\Lambda^2}} \right)^\frac{\gamma_p}{\beta_0}
\end{equation}
where $\beta_0 = 11 - \frac{2}{3} n_f$ and, for $0 \leq p \leq N_{{\rm max}}-1$ (\cite{KM99})~:
\begin{equation}
\label{eqdefgamman}
\gamma_p = \frac{4}{3} \left( 3 + \frac{1}{(p+1)(2p+1)} - 4 \Big( \Psi(2p+2) + \gamma_E \Big) \right)
\end{equation}
where we note $\Psi$ the Digamma function and $\gamma_E$ the Euler-Mascheroni constant.

We use $\Lambda$ = 373~MeV in the $\overline{{\rm MS}}$ scheme with 3 flavours of quarks. We obtained this value after a running of the strong coupling constant computed at four loops \cite{CKS97}, starting from the 2008 world-averaged value of $\alpha_S(M_Z)$ and crossing each quark threshold at (twice) its 2008 averaged mass \cite{PDG08}. This evaluation is in good agreement with a recent textbook one \cite{Ynd99}. The reference scale $Q_0$ has been set to $Q_0^2$ = 3~GeV$^2$.

\subsubsection{Iterative fitting procedure}

In practice we truncate \refeq{eqdefHElecPlusGegenbauer} at some maximum value $N_{{\rm max}}$ of $n$ and we assume the following form for the coefficients $B_{pl}$~:
\begin{equation}
\label{EqtDependenceBpl}
B_{pl}(t,Q^2_0) = \frac{a_{pl}}{1+b_{pl}(t-t_0)^2}
\end{equation}
with $t_0$ a constant and $a_{pl}$ and $b_{pl}$ the free parameters. Their number is $N_{{\rm max}}*(N_{{\rm max}}+3)$. Due to the truncation at $n=N_{{\rm max}}$ the representation of the GPD that we get from a fit of $(a_{pl},b_{pl})$ can hardly be trusted outside the domain of the fit. We take it as a smooth parametrisatin of the data.

The selected JLab data consist in 1001 measurements with $\frac{|t|}{Q^2}<\frac{1}{2}$. We fitted them with $N_{{\rm max}}$ = 2, 3, 4 which corresponds to 10, 18 and 28-parameter fits (performed with Minuit \cite{Jam78}). To constrain the polynomial oscillations in $\xi$, we adopt an interative fitting procedure. We simplify the problem by first working with bins for which we can neglect the $t$-dependence. Indeed, 40~\% of the data with $\frac{|t|}{Q^2}<\frac{1}{2}$   satisfy 0.2~GeV$^2
\leq -t \leq$ 0.4~GeV$^2$. We choose $t_0$ = -0.28~GeV$^2$ in \refeq{EqtDependenceBpl} and extract a first value of the parameters $a_{pl}$. We then add bins with $-t$ between 0.09 and 0.2~GeV$^2$, and between 0.6 and 1.0~GeV$^2$, fitting both $a_{pl}$ and $b_{pl}$, initialising the fit at the values of $a_{pl}$ fitted at the previous iteration, and $b_{pl}$ at 0. We then add the two last $t$-bins to the fit, using at each step the previous extractions of $a_{pl}$ and $b_{pl}$.

\subsubsection{Systematic uncertainties}

The fits with $N_{{\rm max}}$ are qualitatively similar. Their $\chi^2/{\rm d.o.f.}$ are respectively 1.73, 1.61 and 1.78. The comparison of the values of the CFFs derived from these fits gives an estimate of the systematic error on $\mathcal{H}$ induced by the truncation.

Since we assume $H$-dominance, we must take into account the systematic error linked to the neglect of subdominant GPDs. We proceed as in the case of local fits~: we fit the data with the subdominant CFFs set to 0 or to their VGG value, and take the difference as an estimate of the systematic uncertainty.

%%%%%%%%%%%%%%%%%%%%%%%%%%%%%%%%%%%%%%%%%%%%%%%%%%%%%%%%%%%%%%%%%%%%%%%%%%%%%%%%%%%%%%%%%%%%%%%%%%%%
%%%%%%%%%%                                 Troisième partie                               %%%%%%%%%%
%%%%%%%%%%%%%%%%%%%%%%%%%%%%%%%%%%%%%%%%%%%%%%%%%%%%%%%%%%%%%%%%%%%%%%%%%%%%%%%%%%%%%%%%%%%%%%%%%%%%

\section{Results}

\subsection{Extraction of $\ImH$ and $\ReH$}

\begin{figure}
	\begin{center}
		\includegraphics[width=11.5cm]{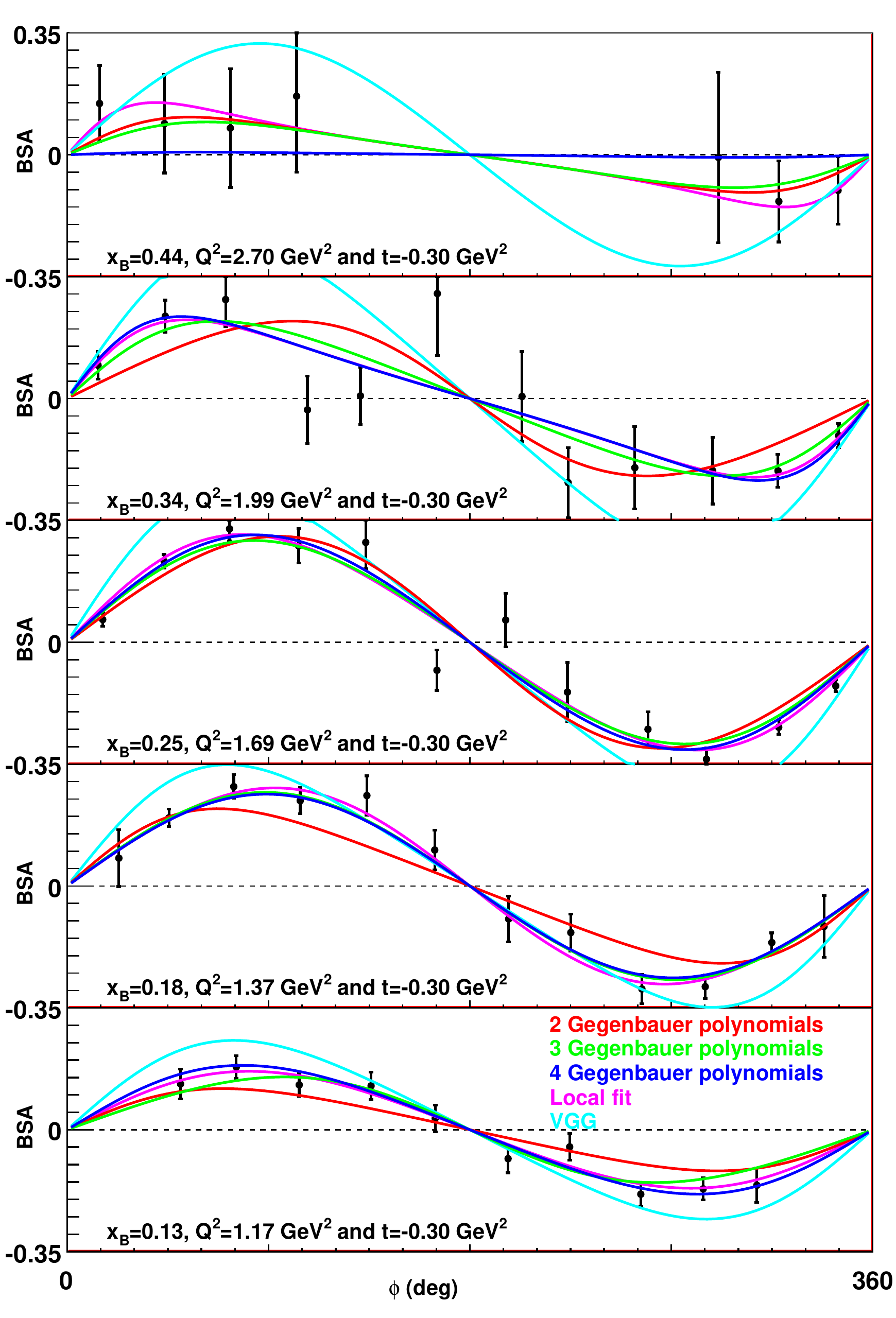}
	\end{center}
	\caption{\label{FigBSATroncature}The $\xb$-range is divided into 5 bins. The BSAs are displayed at the lowest $Q^2$ and for the value of $t$ used at the first iteration of the fit. We plot the results of the global fits with $N_{{\rm max}}$=2, 3, 4, the result of the local fits and the prediction of the VGG code.}
\end{figure}

\begin{figure}
	\begin{center}
		\rotatebox{90}{\includegraphics[width=11.5cm]{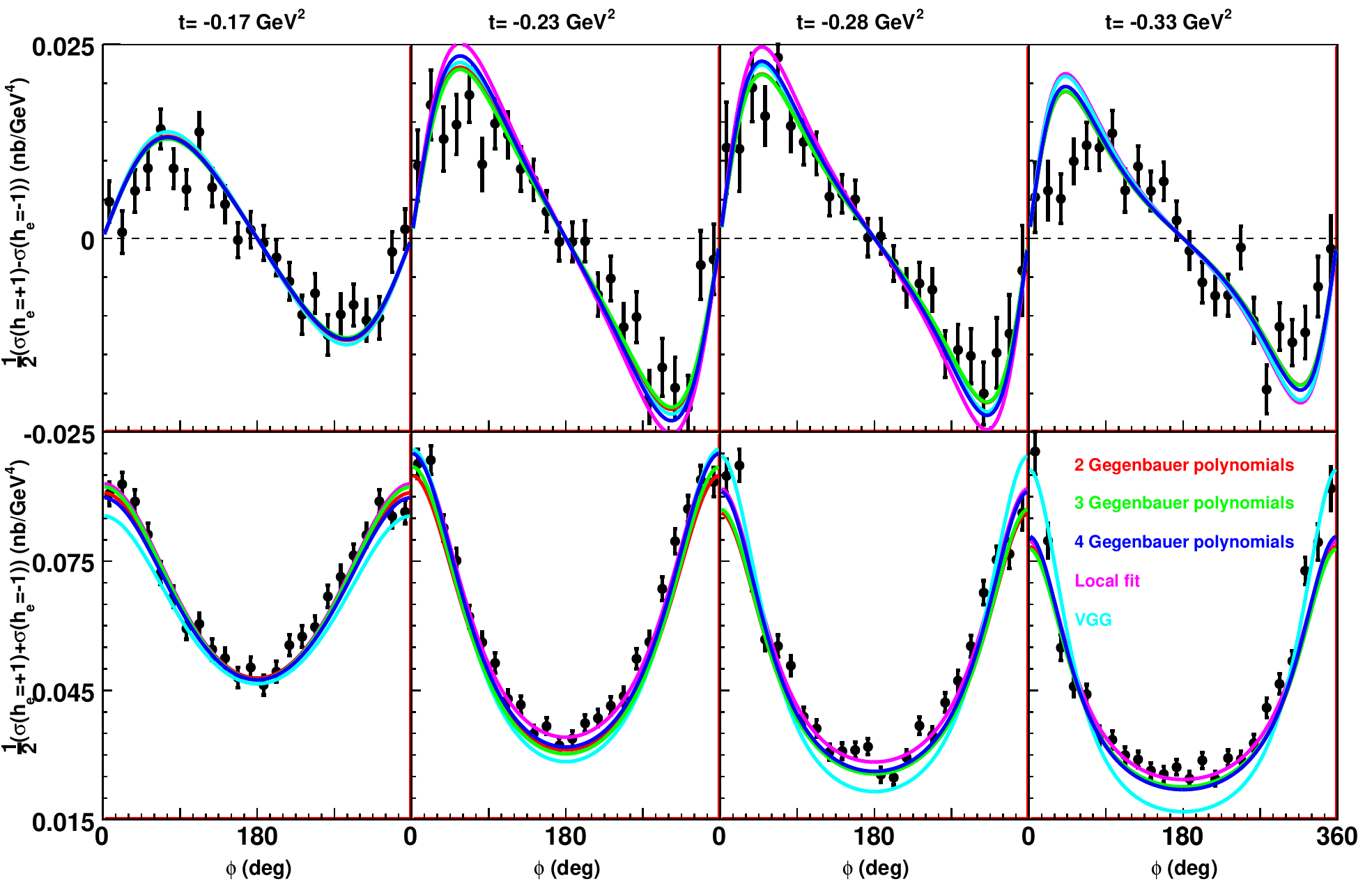}}
	\end{center}
	\caption{\label{FigHallATroncature}Helicity-dependent (up) and helicity-independent (down) cross sections at $\xb$ = 0.36 and $Q^2$ = 2.3~GeV$^2$. We plot the results of the global fits with $N_{{\rm max}}$=2, 3, 4, the result of the local fits and the prediction of the VGG code.}
\end{figure}

The \reffig{FigBSATroncature} and \reffig{FigHallATroncature} display the effect of the truncation of the series in \refeq{eqdefHElecPlusGegenbauer}. When $\xi$ is small, $N_{{\rm max}}$=2 is not enough to describe the BSAs. On the contrary, when $\xi$ is large, we cannot control the fit with $N_{{\rm max}}$=4. This comes from the fact that statistical errors on BSAs are getting larger when $\xi$ grows. The fit with $N_{{\rm max}}$=3 is always good, and close to the local fits, which are optimal by construction. We also see that the VGG model overestimates the data, which is a known feature \cite{Mun06}. It presumably stems from an overestimation of the imaginary parts of CFFs by VGG. At last, global and local fits to helicity-dependent and helicity-independent cross sections are all good, and almost indistinguishable.

We thus choose the fit with $N_{{\rm max}}$=3 as our nominal solution. The systematic uncertainty on $\ImH$ and $\ReH$ linked to the truncation of \refeq{eqdefHElecPlusGegenbauer} is estimated as the maximum of the (absolute values of the) difference between the results for the nominal solution and for the 2 other fits.

\begin{figure}
	\begin{center}
		\includegraphics[width=11.5cm]{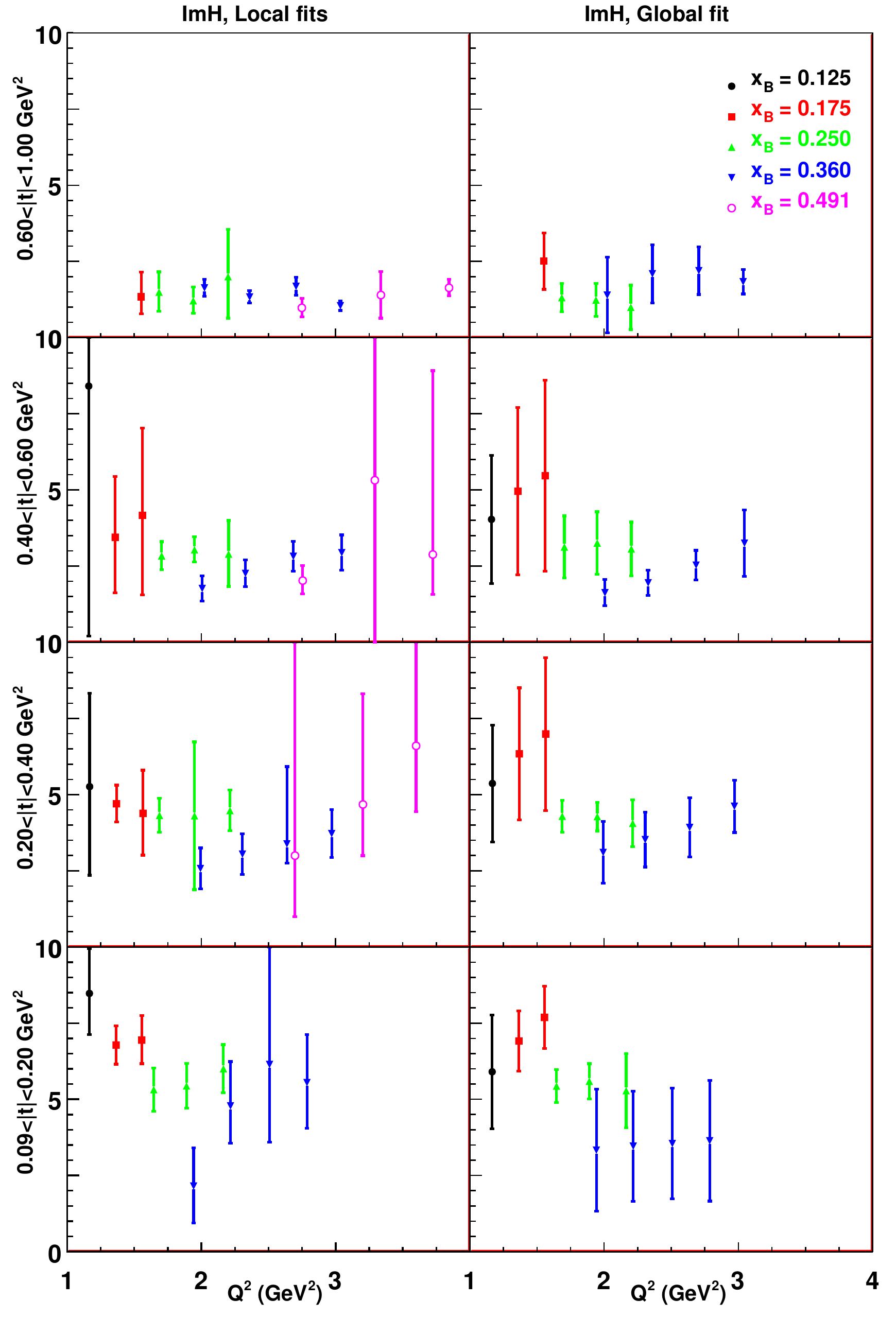}
	\end{center}
	\caption{\label{FigScalingImHHallB}$Q^2$-behaviour of the extracted values of $\ImH$ of local fits (left) and global fit (right) on Hall B kinematics. The error bars include both statistics and systematics.}
\end{figure}

\begin{figure}
	\begin{center}
		\includegraphics[width=11.5cm]{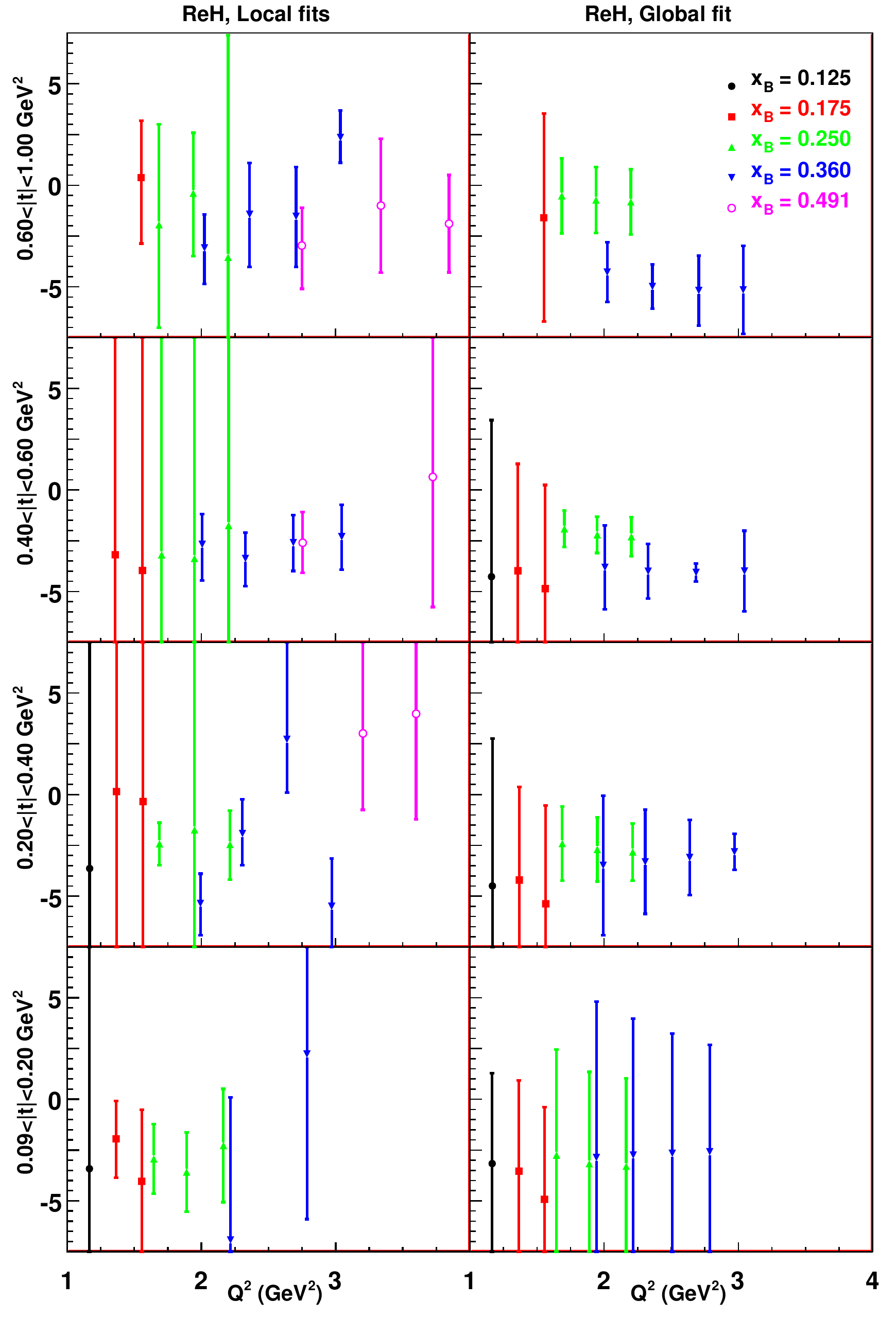}
	\end{center}
	\caption{\label{FigScalingReHHallB}$Q^2$-behaviour of the extracted values of $\ReH$ of local fits (left) and global fit (right) on Hall B kinematics. The error bars include both statistical and systematic uncertainties.}
\end{figure}

\begin{figure}
	\begin{center}
		\includegraphics[width=11.5cm]{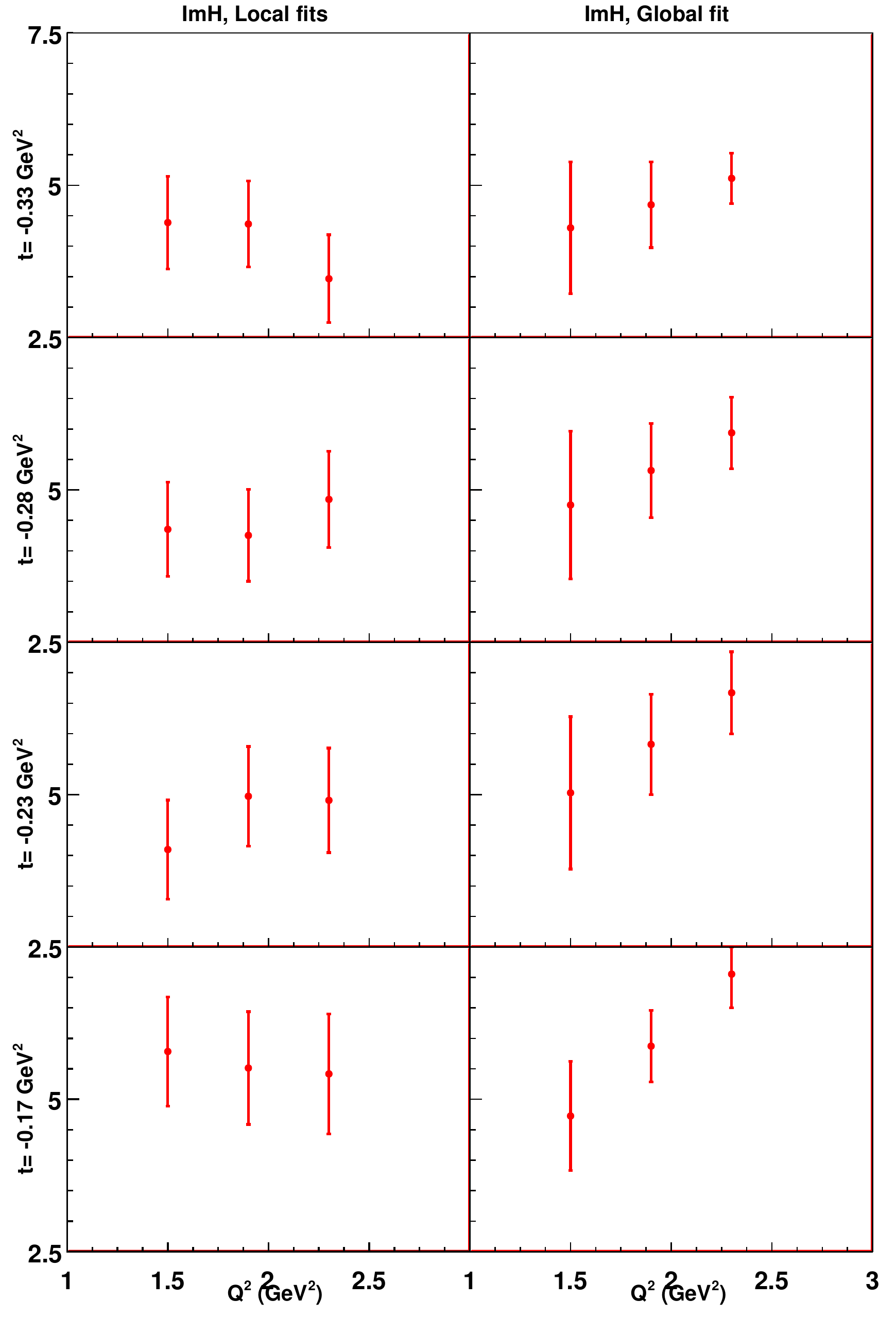}
	\end{center}
	\caption{\label{FigScalingImHHallA}$Q^2$-behaviour of the extracted values of $\ImH$ of local fits (left) and global fit (right) on Hall A kinematics. The error bars include both statistical and systematic uncertainties.}
\end{figure}

\begin{figure}
	\begin{center}
		\rotatebox{90}{\includegraphics[width=11.5cm]{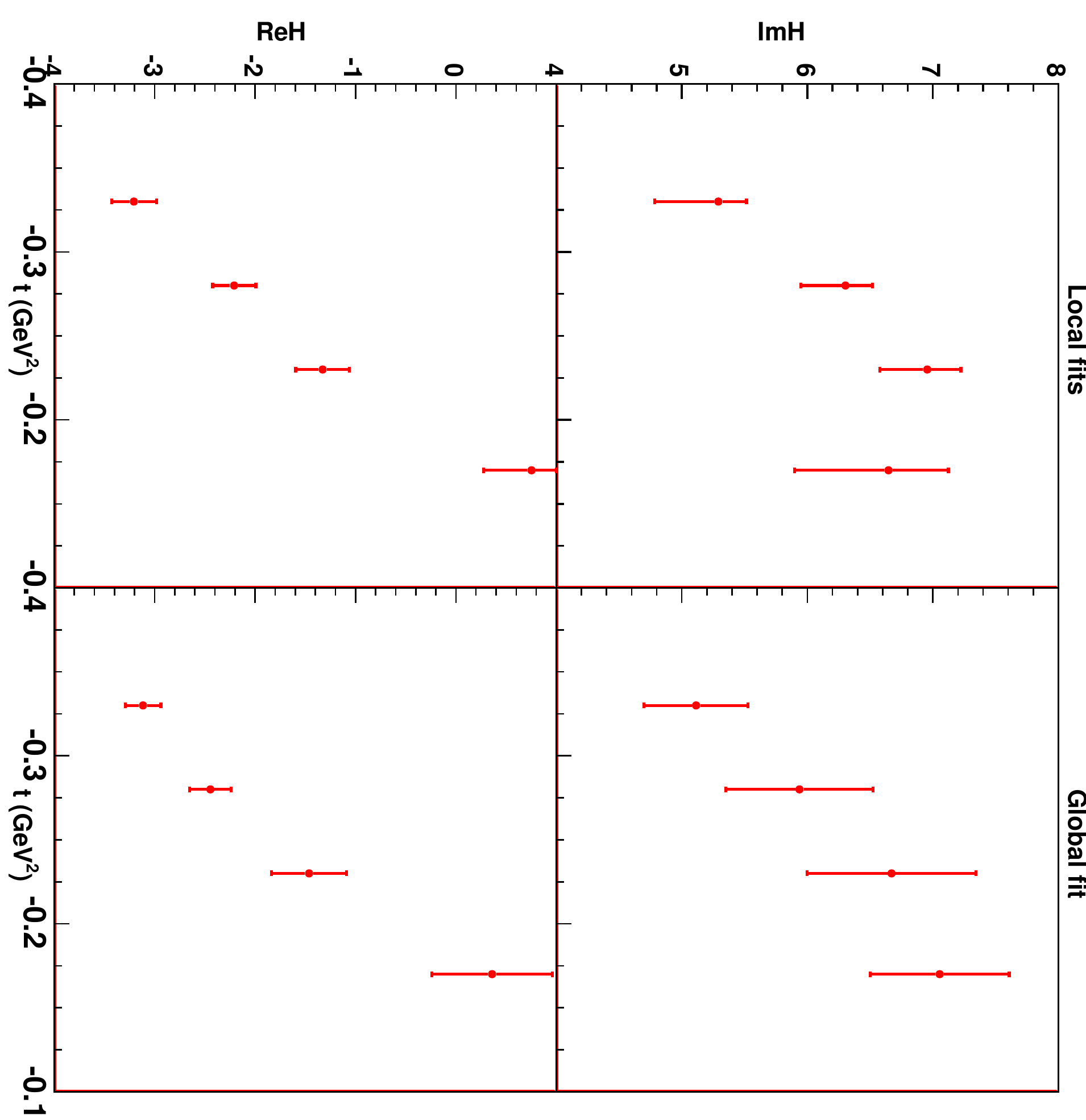}}
	\end{center}
	\caption{\label{FigtbehaviourReHImHHallA}$t$-behaviour of the extracted values of $\ImH$ (up) and $\ReH$ (down) of local fits (left) and global fit (right) on Hall A kinematics ($\xb=0.36$ and $Q^2$ = 2.3~GeV$^{2}$). The error bars include both statistical and systematic uncertainties.}
\end{figure}

The \reffig{FigScalingImHHallB}, \reffig{FigScalingReHHallB}, \reffig{FigScalingImHHallA} and \reffig{FigtbehaviourReHImHHallA} display our results for $\ImH$ and $\ReH$ respectively. Both local fits and global fit give results with comparable accuracy for $\ImH$, but as expected the results of the global fits are smoother. This is especially true concerning $\ReH$~: in this case the local fits suffer from large fluctuations of $\ReH$ with values which fall outside the plot range. However, we could not reliably extract values of the CFFs for the larger values of $\xi$ with the global fit. This is reminiscent of the difficulty in controlling the oscillating behaviour of the polynomial expansion displayed in \reffig{FigBSATroncature}.

The results for local and global fits are almost always compatible, which is a strong consistency check. Both rely on the assumptions of twist 2 accuracy and of $H$-dominance. On one hand, local fits suffer from numerical fluctuations (the 2-parameter local fits are not constrained enough on some bins) but are almost model-independent. On the other hand, global fits are smoother, but suffer from oscillations. That both methods give the same results indicates that fluctuations and oscillations are reasonably controlled in the bins for which results are displayed. Since, in both cases, the total error bars have the same size, we conclude that our estimation of systematic uncertainties due to the truncation of the series \refeq{eqdefHElecPlusGegenbauer} is realistic.

All fits keep data satisfying $\frac{|t|}{Q^2}<\frac{1}{2}$. For local fits, changing the maximal value of $\frac{|t|}{Q^2}$ amounts to dropping points. For global fits, the whole results may be changed,  but the good agreement between the results of both types of fits, and the slow $Q^2$-evolution of the extracted CFFs, indicate that this restricted kinematic region is suitable for an analysis in the GPD framework. Nevertheless, we observe a sizeable scaling deviation on $\ImH$ extracted from Hall A data for $t$ = -0.17~GeV$^{2}$. The choice of the expression of $\xi$ is also related to the issue of scaling. Changing the exact expression of $\xi$ to its asymptotic form $\frac{\xb}{2-\xb}$ induces differences on the extraction of CFFs, the amplitudes of which depend on $\xb$ and $t$. Deviations are noticeable but results with both expressions of $\xi$ are compatible within error bars. This indicates an effect of higher-order power corrections. In view of the whole set of data, the conclusion of early $Q^2$-scaling \cite{Mun06} presumably still holds but a higher-twist study is needed to make it final. 

\reftab{TabResultatHallB} and \reftab{TabResultatHallA} summarize our results. Our error bars are dominated by systematic effects. Typically we obtain a relative accuracy of 20 to 50~\% on $\ImH$, which is quite satisfactory under the assumption of $H$-dominance and given the statistical accuracy of JLab data. On the contrary, $\ReH$ is still largely undetermined, and is never extracted with a precision better than 50~\%. But the imaginary parts of CFF might be the quantities of prime importance as stressed in \cite{Wei09}~: it may be possible to evaluate $\ReH$ (for instance) from the knowledge of $\ImH$ on a wide kinematic range through the use of dispersion relations.

\begin{center}
	\tablefirsthead{
		\hline
		\multicolumn{5}{|c|}{\small\sl Hall B kinematics} \\
		\hline
		$\xb$	& $t~(\gevcarre)$	& $Q^2~(\gevcarre)$ & $\ImH$	& $\ReH$ \\
		\hline
		\hline
	} 
	\tablehead{
		\hline
		\multicolumn{5}{|c|}{\small\sl Hall B kinematics} \\
   		\hline
   		\multicolumn{5}{|r|}{\small\sl continued from previous page~\ldots} \\
   		\hline
		$\xb$	& $t~(\gevcarre)$	& $Q^2~(\gevcarre)$ & $\ImH$	& $\ReH$ \\
		\hline
		\hline
	}
  	\tabletail{
   		\hline
   		\multicolumn{5}{|r|}{\small\sl see following page~\ldots} \\
   		\hline
   	}
   	\tablelasttail{
   		\hline
   		\hline
   	}
	\bottomcaption{
		\label{TabResultatHallB}Global fit extraction of $\ImH$ and $\ReH$ on Hall B kinematics. The horizontal lines gather bins with common $(\xb,t)$ and the horizontal double lines bins with common $t$. The errors are in parenthesis, the first one being statistical, the second and third systematic (respectively truncation and subdominant CFFs).
	}
	\begin{supertabular}{|c|c|c|c|c|}
0.1342 & -0.1337 & 1.1661 & 5.90~(0.01)~(1.57)~(1.01) & -3.16~(0.04)~(1.70)~(4.11) \\ 
\hline
0.1763 & -0.1346 & 1.3651 & 6.91~(0.02)~(0.67)~(0.72) & -3.54~(0.03)~(3.60)~(2.66) \\   
0.1767 & -0.1376 & 1.5557 & 7.69~(0.02)~(0.72)~(0.72) & -4.92~(0.03)~(3.80)~(2.47) \\
\hline  
0.2350 & -0.1465 & 1.6453 & 5.43~(0.03)~(0.34)~(0.42) & -2.75~(0.04)~(5.01)~(1.41) \\   
0.2377 & -0.1448 & 1.8895 & 5.59~(0.03)~(0.35)~(0.46) & -3.17~(0.04)~(4.33)~(1.28) \\
0.2460 & -0.1442 & 2.1641 & 5.28~(0.03)~(1.08)~(0.57) & -3.30~(0.04)~(4.18)~(1.16) \\ 
\hline  
0.3205 & -0.1705 & 1.9424 & 3.33~(0.05)~(1.14)~(1.64) & -2.87~(0.04)~(7.62)~(0.94) \\  
0.3215 & -0.1719 & 2.2170 & 3.46~(0.05)~(0.48)~(1.74) & -2.74~(0.04)~(6.66)~(0.87) \\  
0.3213 & -0.1743 & 2.5078 & 3.55~(0.05)~(0.12)~(1.81) & -2.66~(0.04)~(5.85)~(0.82) \\  
0.3211 & -0.1753 & 2.7865 & 3.64~(0.05)~(0.59)~(1.89) & -2.58~(0.04)~(5.21)~(0.79) \\  
\hline 
\hline 
0.1341 & -0.2840 & 1.1678 & 5.37~(0.01)~(0.60)~(1.82) & -4.50~(0.03)~(4.77)~(5.48) \\ 
\hline  
0.1764 & -0.2798 & 1.3680 & 6.34~(0.01)~(1.52)~(1.54) & -4.21~(0.03)~(3.09)~(3.39) \\  
0.1772 & -0.2819 & 1.5653 & 6.99~(0.01)~(1.97)~(1.55) & -5.37~(0.03)~(3.56)~(3.26) \\  
\hline 
0.2466 & -0.2842 & 1.6881 & 4.29~(0.02)~(0.26)~(0.46) & -2.41~(0.03)~(1.68)~(0.72) \\  
0.2487 & -0.2809 & 1.9490 & 4.28~(0.02)~(0.20)~(0.43) & -2.70~(0.03)~(1.45)~(0.62) \\  
0.2525 & -0.2814 & 2.2131 & 4.06~(0.02)~(0.65)~(0.41) & -2.82~(0.03)~(1.32)~(0.50) \\  
\hline 
0.3399 & -0.3062 & 1.9930 & 3.10~(0.04)~(0.66)~(0.77) & -3.48~(0.03)~(3.43)~(0.06) \\  
0.3431 & -0.3012 & 2.3060 & 3.52~(0.04)~(0.36)~(0.82) & -3.31~(0.03)~(2.56)~(0.02) \\  
0.3447 & -0.2966 & 2.6372 & 3.92~(0.05)~(0.49)~(0.84) & -3.10~(0.03)~(1.84)~(0.09) \\  
0.3480 & -0.2942 & 2.9706 & 4.62~(0.05)~(0.36)~(0.78) & -2.82~(0.03)~(0.87)~(0.15) \\   
\hline 
\hline
0.1331 & -0.4929 & 1.1611 & 4.03~(0.01)~(0.64)~(2.01) & -4.27~(0.02)~(4.81)~(6.03) \\   
\hline
0.1750 & -0.4910 & 1.3580 & 4.96~(0.01)~(1.85)~(2.03) & -3.98~(0.02)~(3.24)~(4.14) \\  
0.1765 & -0.4909 & 1.5611 & 5.47~(0.01)~(2.37)~(2.04) & -4.86~(0.02)~(3.19)~(3.99) \\   
\hline
0.2524 & -0.4875 & 1.7039 & 3.13~(0.02)~(0.57)~(0.84) & -1.91~(0.02)~(0.47)~(0.77) \\  
0.2486 & -0.4873 & 1.9485 & 3.26~(0.02)~(0.54)~(0.87) & -2.21~(0.02)~(0.35)~(0.83) \\  
0.2504 & -0.4883 & 2.2028 & 3.06~(0.02)~(0.36)~(0.81) & -2.30~(0.02)~(0.62)~(0.73) \\  
\hline 
0.3443 & -0.4964 & 2.0062 & 1.63~(0.03)~(0.34)~(0.26) & -3.81~(0.02)~(2.02)~(0.41) \\ 
0.3501 & -0.4938 & 2.3282 & 1.95~(0.04)~(0.37)~(0.19) & -4.00~(0.02)~(1.31)~(0.28) \\  
0.3555 & -0.4889 & 2.6851 & 2.53~(0.04)~(0.48)~(0.02) & -4.06~(0.02)~(0.43)~(0.13) \\  
0.3600 & -0.4854 & 3.0455 & 3.25~(0.04)~(1.06)~(0.25) & -3.99~(0.02)~(1.98)~(0.03) \\  
\hline 
\hline
0.1753 & -0.7741 & 1.5516 & 2.51~(0.01)~(0.68)~(0.63) & -1.60~(0.02)~(4.81)~(1.75) \\  
\hline 
0.2493 & -0.7731 & 1.6847 & 1.31~(0.01)~(0.46)~(0.05) & -0.52~(0.02)~(1.83)~(0.25) \\  
0.2476 & -0.7694 & 1.9394 & 1.24~(0.02)~(0.54)~(0.03) & -0.73~(0.02)~(1.62)~(0.21) \\  
0.2494 & -0.7689 & 2.1990 & 0.99~(0.02)~(0.71)~(0.14) & -0.82~(0.02)~(1.59)~(0.13) \\  
\hline 
0.3516 & -0.7752 & 2.0231 & 1.39~(0.03)~(1.04)~(0.68) & -4.28~(0.02)~(1.45)~(0.26) \\  
0.3597 & -0.7684 & 2.3596 & 2.09~(0.03)~(0.63)~(0.71) & -4.98~(0.02)~(0.97)~(0.49) \\  
0.3607 & -0.7623 & 2.7054 & 2.20~(0.03)~(0.54)~(0.56) & -5.18~(0.02)~(1.60)~(0.66) \\  
0.3582 & -0.7573 & 3.0357 & 1.83~(0.03)~(0.04)~(0.39) & -5.15~(0.02)~(2.04)~(0.76) \\   
\hline 
\hline
	\end{supertabular}		
\end{center}

\begin{table}
	\begin{center}
		\begin{tabular}{|c|c|c|c|c|}
			\hline
			\multicolumn{5}{|c|}{\small\sl Hall A kinematics} \\
   			\hline
			$\xb$	& $t~(\gevcarre)$	& $Q^2~(\gevcarre)$ & $\ImH$	& $\ReH$ \\
			\hline
			\hline
0.3600 & -0.1700 & 1.5000 & 4.73~(0.07)~(0.53)~(0.72) & -1.35~(0.03)~(3.62)~(0.47) \\  
0.3600 & -0.1700 & 1.9000 & 5.87~(0.07)~(0.12)~(0.57) & -0.39~(0.03)~(1.33)~(0.38) \\  
0.3600 & -0.1700 & 2.3000 & 7.05~(0.07)~(0.31)~(0.45) & 0.36~(0.03)~(0.51)~(0.31) \\
\hline  
0.3600 & -0.2300 & 1.5000 & 5.03~(0.06)~(0.94)~(0.82) & -2.77~(0.03)~(2.92)~(0.12) \\  
0.3600 & -0.2300 & 1.9000 & 5.83~(0.06)~(0.36)~(0.74) & -2.04~(0.03)~(1.08)~(0.15) \\  
0.3600 & -0.2300 & 2.3000 & 6.67~(0.06)~(0.19)~(0.64) & -1.46~(0.03)~(0.32)~(0.19) \\ 
\hline 
0.3600 & -0.2800 & 1.5000 & 4.75~(0.05)~(0.99)~(0.69) & -3.45~(0.03)~(2.51)~(0.16) \\  
0.3600 & -0.2800 & 1.9000 & 5.32~(0.05)~(0.43)~(0.64) & -2.90~(0.03)~(0.97)~(0.05) \\  
0.3600 & -0.2800 & 2.3000 & 5.94~(0.05)~(0.20)~(0.55) & -2.45~(0.03)~(0.20)~(0.04) \\
\hline  
0.3600 & -0.3300 & 1.5000 & 4.30~(0.04)~(0.95)~(0.51) & -3.86~(0.03)~(2.20)~(0.40) \\  
0.3600 & -0.3300 & 1.9000 & 4.68~(0.05)~(0.52)~(0.48) & -3.47~(0.03)~(0.91)~(0.24) \\  
0.3600 & -0.3300 & 2.3000 & 5.11~(0.05)~(0.15)~(0.38) & -3.12~(0.03)~(0.15)~(0.10) \\  
		\hline	
		\end{tabular}	
		\caption{\label{TabResultatHallA}Global fit extraction of $\ImH$ and $\ReH$ on Hall A kinematics. The horizontal lines gather bins with common $(\xb,t)$ and the horizontal double lines bins with common $t$. The errors are in parenthesis, the first one being statistical, the second and third systematic (respectively truncation and subdominant CFFs).}
	\end{center} 	
\end{table}

\subsection{Discussion}

\begin{figure}
	\begin{center}
		\rotatebox{90}{\includegraphics[width=11.5cm]{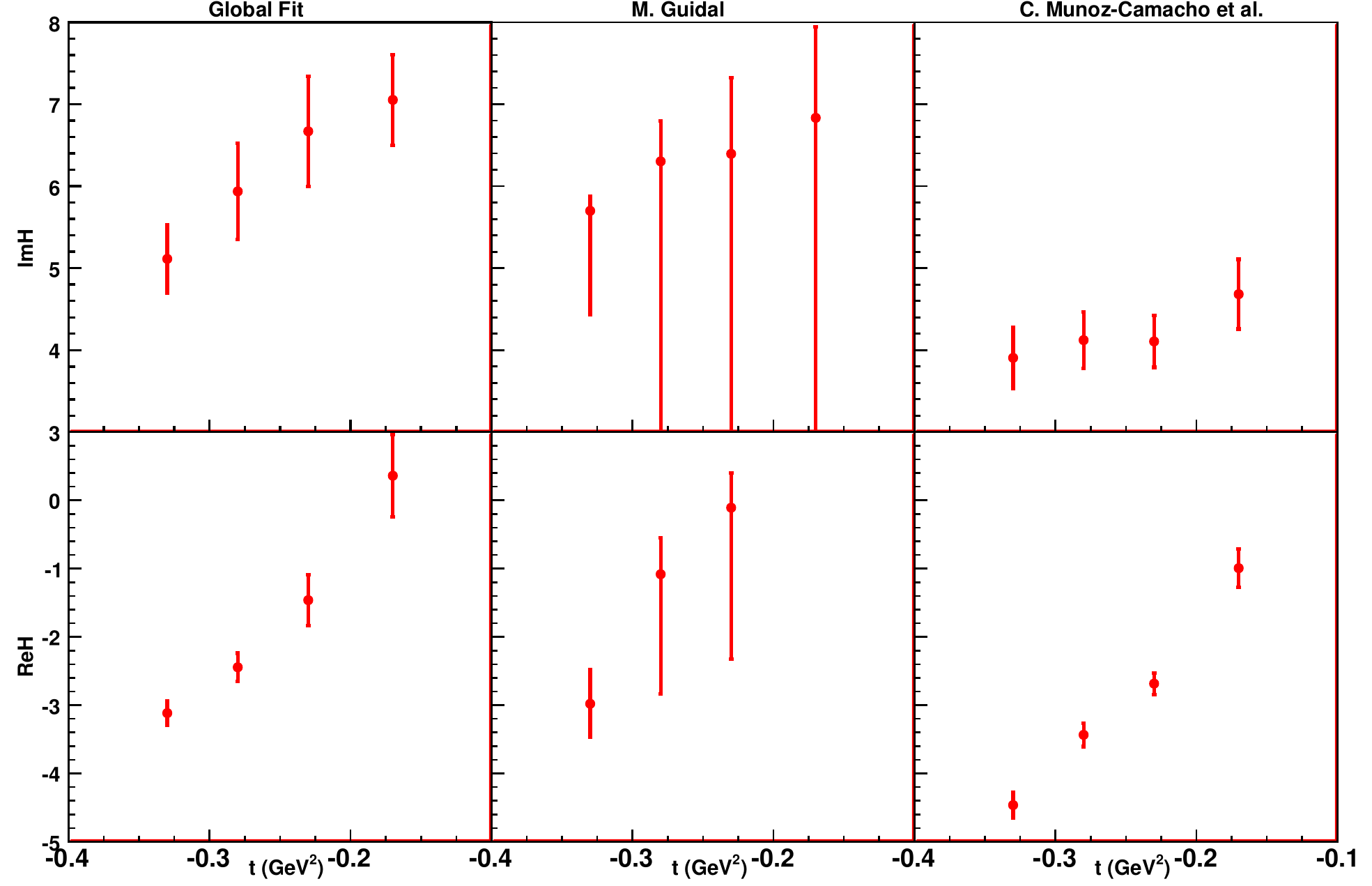}}
	\end{center}
	\caption{\label{FigCompReHImHHallA}$\ReH$ vs $t$ on Hall A kinematics ($\xb$=0.36 and $Q^2$ = 2.3~GeV$^{2}$). We compare our results (left column) to those of M.~Guidal \cite{Gui08} (middle column) and C.~Mu{\~n}oz-Camacho \etal \cite{Mun06} (right column). In the latter column $\mathcal{H} + \frac{\xb}{2-\xb} \left( 1 + \frac{F_2}{F_1} \mathcal{\tilde{H}} \right) - \frac{t}{4 M^2} \frac{F_2}{F_1} \mathcal{E}$ is plot, and not $\mathcal{H}$. The error bars include both statistical and systematic uncertainties.}
\end{figure}

The \reffig{FigCompReHImHHallA} compare our results to a twist 2 model-independent extraction \cite{Gui08} and an extraction with the BMK formalism \cite{Mun06}. Firstly, the use of the GV expressions creates important deviations to the latter extraction. Since the extracted combinations of GPDs are not the same, we will not make the argument more quantitative. Secondly, we obtained results in very good agreement with \cite{Gui08}, but with errors considerably smaller.

\begin{figure}
	\begin{center}
		\rotatebox{90}{\includegraphics[width=11.5cm]{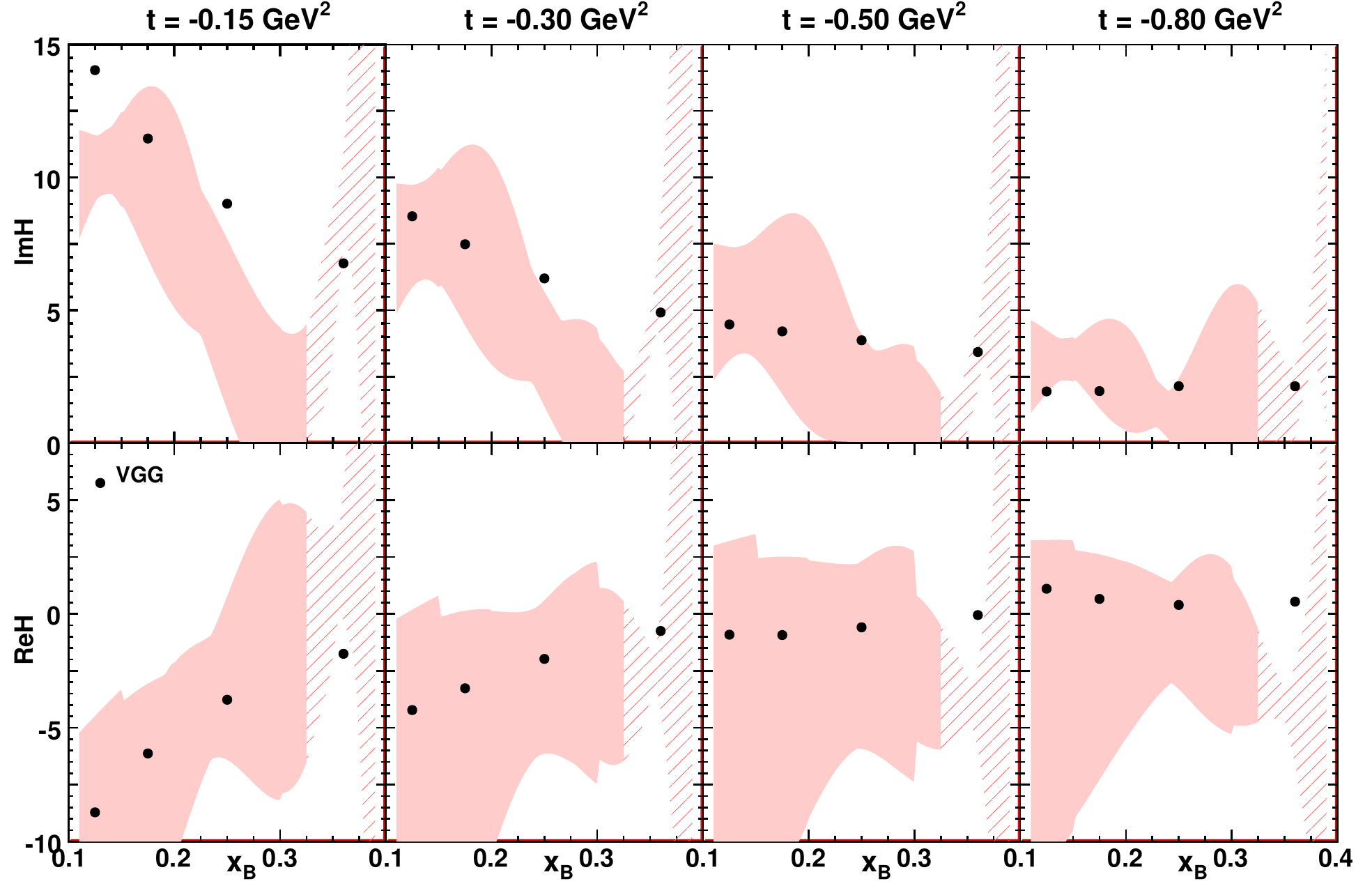}}
	\end{center}
	\caption{\label{FigImHReHCompVGG}Comparisons of the extracted $\ImH$ and $\ReH$ and the predictions of the VGG model. The error bars include both statistical and systematics contributions. Since our errors are dominated by systematics, we can only say that the true value of $\ImH$ and $\ReH$ lie in the colored bands. Moreover, we estimated the systematic errors due to the neglect of subdominant GPDs by averaging over the points where measurements are made, hence enhancing the oscillating behaviour of the fitting curve (see text for more explanations).}
\end{figure}

For a given $t$, and at $Q^2$ = 3~GeV$^2$, the \reffig{FigImHReHCompVGG} displays the $\xb$-dependence of the CFFs we extracted, and compares them to the predictions of the VGG model. To draw the authorised region for $\ImH$ and $\ReH$ we had to estimate the systematic uncertainty induced by the neglect of subdominant CFFs for kinematic configurations for which we have no measurements. According to \reftab{TabResultatHallB} and \reftab{TabResultatHallA}, this error weakly depends on $Q^2$ for fixed $t$ and $\xb$. We averaged it over each $(\xb,t)$-bin, and used it to draw the color band of \reffig{FigImHReHCompVGG}. Doing so, we enhanced the oscillating behaviour of the contours of the plot by adding discontinuities. Since the errors are mostly due to systematic effects, the points in the colored domain have the same probability. 

This being stated, we notice that the extracted values of $\ImH$ have an $\xb$-dependence similar to that predicted by the VGG model. We again observe that VGG tends to overestimate $\ImH$. The extracted CFFs are often but not always compatible with VGG. The errors on the extracted $\ReH$ and $\ImH$ in the hatched zone may be underestimated. We already noticed that polynomial oscillations are more severe at large $\xb$, where the data are less accurate, as shown in \reffig{FigBSATroncature} and \reffig{FigHallATroncature}. However, from \reffig{FigScalingImHHallB} and \reffig{FigScalingReHHallB} we know that the extraction is independent of the type (local or global) of the fit for $0.3 \leq \xb \leq 0.4$. Thus these data should not be rejected but it is necessary to stay cautious when interpreting them.

%%%%%%%%%%%%%%%%%%%%%%%%%%%%%%%%%%%%%%%%%%%%%%%%%%%%%%%%%%%%%%%%%%%%%%%%%%%%%%%%%%%%%%%%%%%%%%%%%%%%
%%%%%%%%%%                                   Conclusions                                  %%%%%%%%%%
%%%%%%%%%%%%%%%%%%%%%%%%%%%%%%%%%%%%%%%%%%%%%%%%%%%%%%%%%%%%%%%%%%%%%%%%%%%%%%%%%%%%%%%%%%%%%%%%%%%%

\section*{Conclusions}

Working at leading twist, and assuming $H$-dominance, we extracted $\ImH$ and $\ReH$ with two different methods. The local fits do not benefit from the wide kinematic coverage of Hall B data, and are bound to produce rather large statistical uncertainties when fitting two independent parameters from an asymmetry measured at (at most) twelve $\phi$-bins. On the opposite, the global fit dramatically decreases the systematic uncertainties and uses the kinematic coverage to disentangle the contributions of the different CFFs at the expense of a fitting Ansatz. The good agreement between the results of both extractions is a strong consistency check. It demonstrates that the extraction of $\ImH$ and $\ReH$ and the estimation of the systematics related to this fitting procedure produce relatively model-independent results, at least for $\xb$ not too large. This estimation of the systematics will be refined but these first results are already encouraging. One immediate advantage is the smoothness of the results. This comparison and the weak $Q^2$-dependency of the results also validates \textit{a posteriori} the restriction to kinematic configurations with $\frac{-t}{Q^2}<\frac{1}{2}$.

It is one of the first global fits of measurements from different experiments. In spite of the difficulty of controlling polynomial oscillations, the (fitted) behaviour of $\ImH$ and $\ReH$ versus $\xb$ is similar to the VGG model prediction. The $t$-dependence of our global fit is also in fair agreeement with that obtained in the first GPD analysis of Hall A data in the BMK framework. As expected, we find that the VGG model tends to overestimate the physical value of $\ImH$.

Our results are dominated by systematic uncertainties. Their origin is twofold. Firstly, we assumed $H$-dominance and neglected the contribution of $E$, $\tilde{E}$ and $\tilde{H}$. Secondly, the Gegenbauer and Legendre expansions induce oscillations in the partial sums of the series. We may hope to reduce these systematic uncertainties in the near future using additional BSA measurements \cite{Ung09} (unpolarised proton target) and \cite{Sed09} (longitudinaly polarised proton target) which will put stronger constraints on the global fits. The extension of our procedure to smaller values of $\xb$ is in progress.

However, our total errors are already of reasonable size, since they are comparable to, or smaller than those coming from previous extractions. We typically obtain a 20 to 50~\% accuracy on $\ImH$, which is already good regarding our hypothesis of $H$ dominance.  

At last, C.~Mu{\~n}oz-Camacho \etal \cite{Mun06} performed the extraction of a combination of GPDs with the BMK formulae. They concluded that Hall A data indicates twist 2 dominance of DVCS through early $Q^2$-scaling. Using the new GV formalism, we tend to come to the same conclusion in this study, but we plan to work out a refined analysis at twist 3 in the Wandura-Wilczek approximation to make sure we distinguish between power law (higher twist) and logarithmic (evolution) behaviours.

%%%%%%%%%%%%%%%%%%%%%%%%%%%%%%%%%%%%%%%%%%%%%%%%%%%%%%%%%%%%%%%%%%%%%%%%%%%%%%%%%%%%%%%%%%%%%%%%%%%%
%%%%%%%%%%                     Papier de K. Kumericki et D. Müller                        %%%%%%%%%%
%%%%%%%%%%%%%%%%%%%%%%%%%%%%%%%%%%%%%%%%%%%%%%%%%%%%%%%%%%%%%%%%%%%%%%%%%%%%%%%%%%%%%%%%%%%%%%%%%%%%

\paragraph*{Note added} During the writing of this paper, K.~Kumericki and D.~Müller released a detailed model-dependent fit of Hermes and JLab DVCS measurements. We will compare their results to ours (both present results and the on-going extension to smaller $\xb$) in a future study.

%%%%%%%%%%%%%%%%%%%%%%%%%%%%%%%%%%%%%%%%%%%%%%%%%%%%%%%%%%%%%%%%%%%%%%%%%%%%%%%%%%%%%%%%%%%%%%%%%%%%
%%%%%%%%%%                                  Remerciements                                 %%%%%%%%%%
%%%%%%%%%%%%%%%%%%%%%%%%%%%%%%%%%%%%%%%%%%%%%%%%%%%%%%%%%%%%%%%%%%%%%%%%%%%%%%%%%%%%%%%%%%%%%%%%%%%%

\section*{Acknowlegments}

The author would like to thank J.~Ball, M.~El Yakoubi, M.~Garçon, F.-X.~Girod, P.A.M.~Guichon, M.~Guidal, P.~Konczykowki, C.~Mu{\~n}oz-Camacho, S.~Procureur and F.~Sabatié for many fruitfull and stimulating discussions, M.~Guidal for providing VGG values of $\ReH$ and $\ImH$ for JLab Hall B kinematics, and P.A.M.~Guichon and M.~Vanderhaeghen for allowing an early use of the results of their study on DVCS on proton at twist 2.

This work was supported in part by the Commissariat à l'Energie Atomique and the GDR n° 3034 Physique du Nucleon.

%%%%%%%%%%%%%%%%%%%%%%%%%%%%%%%%%%%%%%%%%%%%%%%%%%%%%%%%%%%%%%%%%%%%%%%%%%%%%%%%%%%%%%%%%%%%%%%%%%%%
%%%%%%%%%%                                  Bibliographie                                 %%%%%%%%%%
%%%%%%%%%%%%%%%%%%%%%%%%%%%%%%%%%%%%%%%%%%%%%%%%%%%%%%%%%%%%%%%%%%%%%%%%%%%%%%%%%%%%%%%%%%%%%%%%%%%%


\begin{thebibliography}{2}
	\bibitem{Ji96} X.~Ji, \emph{Gauge-invariant decomposition of nucleon spin and its spin-off}, Phys.~Rev.~Lett.~78~(1997)~610, arXiv:hep-ph/9603249.
	\bibitem{DGPR97} M.~Diehl, T.~Gousset, B.~Pire and J.P.~Ralston, \emph{Testing the handbag contribution to exclusive virtual Compton scattering}, Phys.~Lett.~B411~(1997)~193, arXiv:hep-ph/9706344.
	\bibitem{GPV01} K.~Goeke, M.V.~Polyakov and M.~Vanderhaeghen, \emph{Hard exclusive reactions and the structure of hadrons}, Prog.~Part.~Nucl.~Phys.~47~(2001)~401, arXiv:hep-ph/0106012.
	\bibitem{Die03} M.~Diehl, \emph{Generalized parton distributions}, Phys.~Rept.~388~(2003)~41, arXiv:hep-ph/0307382.
	\bibitem{BR05} A.V.~Belitsky and A.V.~Radyushkin, \emph{Unraveling hadron structure with generalized parton distributions}, Phys.~Rept.~418~(2005)~1, arXiv:hep-ph/0504030.
	\bibitem{BP07} S.~Boffi and B.~Pasquini, \emph{Generalized parton distributions and the structure of the nucleon}, Riv.~Nuovo~Cim.~30~(2007)~387, arXiv:0711.2625 [hep-ph].
	\bibitem{Air01} A.~Airapetian \etal (HERMES Coll.), \emph{easurement of the beam spin azimuthal asymmetry associated with deeply virtual Compton scattering}, Phys.~Rev.~Lett.~87~(2001)~182001, arXiv:hep-ex/0106068.	
	\bibitem{Adl01} C.~Adloff \etal (H1 Coll.), \emph{Measurement of deeply virtual Compton scattering at HERA}, Phys.~Lett.~B517~(2001)~47, arXiv:hep-ex/0107005.	
	\bibitem{Ste01} S.~Stepanyan \etal (CLAS Coll.), \emph{First observation of exclusive deeply virtual Compton scattering in polarized electron beam asymmetry measurements}, Phys.~Rev.~Lett.~87~(2001)~182002, arXiv:hep-ex/0107043.	
	\bibitem{Che03} S.~Chekanov el al. (ZEUS Coll.), \emph{Measurement of deeply virtual Compton scattering at HERA}, Phys.~Lett.~B573~(2003)~46, arXiv:hep-ex/0305028.	
	\bibitem{Che06} S.~Chen \etal (CLAS Coll.), \emph{Measurement of Deeply Virtual Compton Scattering with a polarized proton target}, Phys.~Rev.~Lett.~97~(2006)~072002, arXiv:hep-ex/0605012.	
	\bibitem{Mun06} C.~Mu{\~n}oz-Camacho \etal, \emph{Scaling tests of the cross-section for deeply virtual compton scattering}, Phys.~Rev.~Lett.~97~(2006)~262002, arXiv:nucl-ex/0607029v2.
	\bibitem{Aar07} F.D.~Aaron, \emph{Measurement of deeply virtual Compton scattering and its t-dependence at HERA}, Phys.~Lett.~B659~(2008)~796, arXiv:0709.4114 [hep-ex].	
	\bibitem{Gir07} F.-X.~Girod \etal, \emph{Deeply virtual Compton scattering beam-spin asymmetries}, Phys.~Rev.~Lett.~100~(2008)~162002, arXiv:0711.4805 [hep-ex].		
	\bibitem{Air08} A.~Airapetian \etal (HERMES Coll.), \emph{Measurement of Azimuthal Asymmetries With Respect To Both Beam Charge and Transverse Target Polarization in Exclusive Electroproduction of Real Photons}, JHEP~0806~(2008)~066, arXiv:0802.2499 [hep-ex].	
	\bibitem{BMK02} A.V.~Belitsky, D.~Müller and A.~Kirchner, \emph{Theory of deeply virtual Compton scattering on the nucleon}, Nucl.~Phys.~B629~(2002)~323, arXiv:hep-ph/0112108v2.
	\bibitem{BM08} A.V.~Belitsky and D.~Müller, \emph{Refined analysis of photon leptoproduction off spinless target}, arXiv:0809.2890 [hep-ph].
	\bibitem{GV08} P.A.M.~Guichon and M.~Vanderhaegen, \emph{Analytic ee'$\gamma$ cross section}, \url{http://lpsc.in2p3.fr/Indico/getFile.py/access?contribId=1&resId=0&materialId=slides&confId=206}, private communication.
	\bibitem{Ter05} O.V.~Teryaev, \emph{Analytic properties of hard exclusive amplitudes}, Contribution to 11th International Conference on Elastic and Diffractive Scattering: Towards High Energy Frontiers: The 20th Anniversary of the Blois Workshops, Chateau de Blois, Blois, France, 15-20 May 2005, arXiv:hep-ph/0510031.
	\bibitem{AT07} I.V.~Anikin and O.V.~Teryaev, \emph{Dispersion relations and QCD factorization in hard reactions}, Talk given at International Conference on Hadron Structure (HS 07), Modra-Harmonia, Slovakia, 3-7 Sep 2007, arXiv:0710.4211 [hep-ph].
	\bibitem{DI07} M.~Diehl and D.Y.~Ivanov, \emph{Dispersion representations for hard exclusive processes: beyond the Born approximation}, Eur.~Phys.~J.~C.~52~(2007)~919, arXiv:0707.0351 [hep-ph].
	\bibitem{PW99} M.V.~Polyakov and C.~Weiss, \emph{Skewed and double distributions in the pion and the nucleon}, Phys.~Rev.~D~60~(1999)~114017, arXiv:hep-ph/9902451.
	\bibitem{VG98} M.~Vanderhaeghen and P.A.M.~Guichon, \emph{Virtual Compton Scattering off the nucleon}, Prog.~Part.~Nucl.~Phys.~41~(1998)~125, arXiv:hep-ph/9806305.
	\bibitem{Bac04} A.~Bacchetta \etal, \emph{Single-spin asymmetries : the Trento conventions}, Phys.~Rev.~D70~(2004)~117504, hep-ph/0410050. 
	\bibitem{VGG98} M.~Vanderhaeghen, P.A.M.~Guichon, M.~Guidal, \emph{Hard electroproduction of photons and mesons on the nucleon}, Phys.~Rev.~Lett.~80~(1998)~5064.
	\bibitem{VGG99} M.~Vanderhaeghen, P.A.M.~Guichon, M.~Guidal, \emph{Deeply virtual electroproduction of photons and mesons on the nucleon: Leading order amplitudes and power corrections}, Phys.~Rev.~D60~(1999)~094017.
	\bibitem{GPRV05} M.~Guidal, M.V.~Polyakov, A.V.~Radyushkin and M.~Vanderhaeghen, \emph{Nucleon form-factors from generalized parton distributions}, Phys.~Rev.~D72~(2005)~054013, arXiv:hep-ph/0410251.
	\bibitem{Gui08} M.~Guidal, \emph{A fitter code for Deep Virtual Compton Scattering and Generalized Parton Distributions}, Eur.~Phys~.~J.~A37~(2008)~319, arXiv:0807.2355 [hep-ph].
	\bibitem{CF98} J.C.~Collins and A.~Freund, \emph{Proof of factorization for deeply virtual Compton scattering in QCD}, Phys.~Rev.~D59~(1999)~074009, arXiv:her-ph/9801262.
	\bibitem{PSTS08} M.V.~Polyakov et K.~Semenov-Tian-Shansky, \emph{Dual parametrization of GPDs versus double distribution Ansatz}, arXiv:0811.2901 [hep-ph].
	\bibitem{PS02} M.V.~Polyakov et A.G.~Shuvaev, \emph{On 'dual' parametrizations of generalized parton distributions}, arXiv:hep-ph/0207153v1. 
	\bibitem{GT06} V.~Guzey et T.~Teckentrup, \emph{Dual parametrization of the proton generalized parton distribution $H$ and $E$, and description of the deeply virtual Compton scattering cross sections and asymmetries}, arXiv:hep-ph/0607099v1.
	\bibitem{GT08} V.~Guzey et T.~Teckentrup, \emph{On the mistake in the implementation of the minimal model of the dual parameterization and resulting inability to describe the high-energy DVCS data}, Phys.~Rev.~D79~(2009)~017501, arXiv:0810.3899 [hep-ph].
	\bibitem{KM99} N.~Kivel et L.~Mankiewicz, \emph{Conformal string operators and evolution of skewed parton distributions}, arXiv:hep-ph/9903531v1.
	\bibitem{CKS97} K.G. Chetyrkin, B.A.~Kniehl and M.~Steinhauser, \emph{Decoupling relations to O($\alpha_S^3$) and their connection to low-energy theorems}, Nucl.~Phys.~B~510~(1998)~61, arXiv:hep-ph/9708255.
	\bibitem{PDG08} Particle Data Group, \url{http://pdg.lbl.gov/index.html}.
	\bibitem{Ynd99} F.J.~Yndurain, \emph{The theory of quark and gluon interactions}, Springer-Verlag 1999, third edition.
	\bibitem{Jam78} F.~James, \emph{Minuit, Function minimisation and error analysis, reference manual, v94.1}, CERN D506, august 1998 edition.
	\bibitem{BGMS97} A.V.~Belitsky, B.~Geyer, D.~Müller and A.~Schäfer, \emph{On the leading logarithmic evolution of the off forward distributions}, Phys.~Lett.~B421~(1998)~31, arXiv:hep-ph/9710427.
	\bibitem{Wei09} C.~Weiss, \emph{Generalized parton distributions: Status and perspectives}, in Proceedings of SPIN2008, University of Virginia, October 6-11, 2008, arXiv:0902.2018 [hep-ph].
	\bibitem{Ung09} M.~Ungaro, \emph{E1-DVCS (2) run status}, CLAS12 European Meeting, February 25-28, 2009- Genova, Italy, \url{http://www.ge.infn.it/~clas12/clas12/talks/friday_deep/e1-dvcs\%20stattus.pptx}.
	\bibitem{Sed09} E.~Seder, \emph{Status Update on the eg1-dvcs Experiment}, CLAS12 European Meeting, February 25-28, 2009- Genova, Italy, \url{http://www.ge.infn.it/~clas12/clas12/talks/friday_deep/seder.ppt}.
	\bibitem{KM09} K.~Kumericki and D.~Müller, \emph{Deeply virtual Compton scattering at small x$_B$ and the access to the GPD $H$}, arXiv:0904.0458 [hep-ph].
\end{thebibliography}
\end{document}